\documentclass[runningheads,a4paper]{llncs}

\usepackage{latexsym}
\usepackage{amssymb}
\usepackage{amsmath}
\usepackage{alltt}
\usepackage[all]{xy}

\usepackage{graphicx}

{\refstepcounter{defcounter} \begin{trivlist} \item[]%
\theoremstart{Algorithm}{\thedefcounter}}%
{\end{trivlist}}

\def\defemb#1#2{\expandafter\def\csname #1\endcsname
                              {\relax\ifmmode #2\else\hbox{$#2$}\fi}}
\defemb{cA}{{\cal A}}
\defemb{cD}{{\cal D}}
\defemb{cG}{{\cal G}}
\defemb{cH}{{\cal H}}
\defemb{cQ}{{\cal Q}}
\defemb{cR}{{\cal R}}
\defemb{cT}{{\cal T}}
\defemb{cV}{{\cal V}}

\newcommand{\mgg}{\mathit{mgg}}

\newcommand{\sgeq}{\geqslant}

\newcommand{\snorm}{||\cdot||}
\newcommand{\conc}{\bullet}

\newcommand{\ignore}[1]{} 

\begin{document}

\title{Improving Size-Change Analysis in\\ Offline Partial
  Evaluation\thanks{ This work has been partially supported by the EU
    (FEDER) and the Spanish MEC/MICINN under grants
    TIN2005-09207-C03-02, TIN2008-06622-C03-02, and \emph{Acci\'on
      Integrada} HA2006-0008} }

\author{Michael Leuschel\inst{1} \and Salvador Tamarit\inst{2} \and
  Germ\'an Vidal\inst{2}}

\institute{ Institut f{\"u}r Informatik, Universit{\"a}t
  D{\"u}sseldorf,
  D-40225, D{\"u}sseldorf, Germany\\
  \email{leuschel@cs.uni-duesseldorf.de}\\
  \and DSIC, Technical University of Valencia,
  E-46022, Valencia, Spain\\
  \email{$\{$stamarit,gvidal$\}$@dsic.upv.es} }


\tocauthor{Michael Leuschel (University of D\"{u}sseldorf), Salvador
  Tamarit (Technical University of Valencia), German Vidal (Technical
  University of Valencia)}


\maketitle

\setcounter{page}{44}

\begin{abstract}
  Some recent approaches for scalable offline partial evaluation of
  logic programs include a size-change analysis for ensuring both so
  called local and global termination. In this work|inspired by
  experimental evaluation|we introduce several improvements that may
  increase the accuracy of the analysis and, thus, the quality of the
  associated specialized programs.  We aim to achieve this while
  maintaining the same complexity and scalability of the recent works.
\end{abstract}

\section{Introduction}

Partial evaluation \cite{JGS93} is a well-known technique for program
specialization. In this work, we consider the so called \emph{offline}
approach, which consists of two clearly separated phases: binding-time
analysis and proper specialization. Basically, the binding-time
analysis should annotate the source code in order to drive the
specialization process. Roughly speaking,
\begin{itemize}
\item every atom is annotated as either \textsf{unfold} (the atom
  can be unfolded) or \textsf{memo} (the atom should not be unfolded),
  and
\item every predicate's argument is classified as either
  \textsf{static} (definitely known at specialization time) or
  \textsf{dynamic} (possibly unknown at specialization time).
\end{itemize}
We say that the annotations are \emph{safe} if static arguments are
actually ground at specialization time and the termination of the
specialization is ensured. Termination issues are usually classified
into local and global termination:
\begin{itemize}
\item local termination ensures that no atom is infinitely unfolded;
\item global termination guarantees that only finitely many atoms are
  specialized (i.e., that we do not create infinite specializations of
  the same predicate).
\end{itemize}
The main component of a binding-time analysis is a termination
analysis that allows us to guarantee both local and global termination
of the specialization process. In \cite{Vid07}, a \emph{strong}
termination analysis|based on the so called size-change termination
principle \cite{LJB01}|for logic programs is introduced. Strong
termination means termination w.r.t.\ all selection rules. Although
this is a rather strong condition, it allows us to design much faster
binding-time analysis (see \cite{LV08}). 

In this paper, we identify several weaknesses of the original
size-change analysis of \cite{Vid07} and present different proposals
that improve the accuracy of the specialization process.

\section{Size-Change Termination Analysis} \label{scys}

In this section, we informally present the basis of the
quasi-termination analysis for logic programs of \cite{Vid07}.

We say that a query $Q$ is \emph{strongly terminating} w.r.t.\ a
program $P$ if every SLD derivation for $Q$ with $P$ is finite.  We
denote by $calls_P^\cR(Q_0)$ the set of calls in the computations of a
goal $Q_0$ within a logic program $P$ and a computation rule $\cR$.
The query $Q$ is \emph{strongly quasi-terminating} if, for every
computation rule $\cR$, the set $call_P^\cR(Q)$ contains finitely many
nonvariant atoms.  A program $P$ is strongly (quasi-)terminating
w.r.t.\ a set of queries $\cQ$ if every $Q \in \cQ$ is strongly
(quasi-)terminating w.r.t.\ $P$.
For conciseness, in the remainder of this paper, we write
``(quasi-)termination'' to refer to ``strong (quasi-)termination.''

Size-change analysis is based on constructing graphs that represent
the decrease of the arguments of a predicate from one call to another.
For this purpose, some ordering on terms is required. 

\begin{definition}[reduction pair] 
  We say that $(\succsim,\succ)$ is a reduction pair 
  if $\;\succsim$ is a quasi-order and $\succ$ is a well-founded order
  where both $\succsim$ and $\succ$ are closed under substitutions and
  compatible (i.e., $\succsim \circ \succ \;\subseteq\; \succ$ and
  $\succ \circ \succsim \;\subseteq\; \succ$ but $\succsim
  \;\subseteq\; \succ$ is not necessary).
\end{definition}
In logic programming, however, termination analyses usually rely on
the use of \emph{norms} which measure the size of terms.  In
\cite{Vid07}, {reduction orders} $(\succsim,\succ)$ \emph{induced}
from symbolic norms $\snorm$ are used:

\begin{definition}[symbolic norm \cite{DLSS01,LS97}] \label{snorm-def}
  Given a term $t$,
  \[
  ||t|| = \left\{
  \begin{array}{ll}
  m + \sum_{i=1}^{n} k_i ||t_i|| & \mbox{ if } t = f(t_1,\ldots,t_n),~n\sgeq 0\\
  t & \mbox{ if $t$ is a variable }
  \end{array}
  \right.
  \]
  where $m$ and $k_1,\ldots,k_n$ are non-negative integer constants
  depending only on $f/n$. Note that we associate a variable over
  integers with each logical variable (we use the same name for both
  since the meaning is clear from the context).
\end{definition}
The introduction of variables in the range of the norm provides a
simple mechanism to express dependencies between the sizes of terms.

The associated induced orders $(\succsim,\succ)$ are defined as
follows: $t_1 \succ t_2$ (respec.\ $t_1 \succsim t_2$) if
$||t_1\sigma|| > ||t_2\sigma||$ (respec.\ $||t_1\sigma|| \sgeq
||t_2\sigma||$) for all substitution $\sigma$ that makes
$||t_1\sigma||$ and $||t_2\sigma||$ ground (e.g., an integer
constant).
Two popular instances of symbolic norms are the symbolic
\emph{term-size} norm $\snorm_{ts}$ (which sums the arities of the
term symbols) and the symbolic \emph{list-length norm} $\snorm_{ll}$
(which counts the number of elements of a list), e.g.,
\[
\begin{array}{rcl}
  f(X,Y,a) \succ_{ts} f(X,a,b) & \mbox{since} & 
     ||f(X,Y,a)||_{ts} = X+Y+3 > X+3 = ||f(X,a,b)||_{ts} \\\protect
  [X|R] \succsim_{ll} [s(X)|R] & \mbox{since} & 
     ||[X|R]||_{ll} = R+1 \sgeq R+1 = ||[s(x)|R]||_{ll} \\
\end{array}
\]
Now, we produce a \emph{size-change graph} $\cG$ for every pair
$(H,B_i)$ of every clause $H \leftarrow B_1,\ldots,B_n$ of the
program, with edges between the arguments of $H$ and $B_i$ when the
size of the corresponding terms decrease w.r.t.\ a given reduction
pair $(\succsim,\succ)$. 

  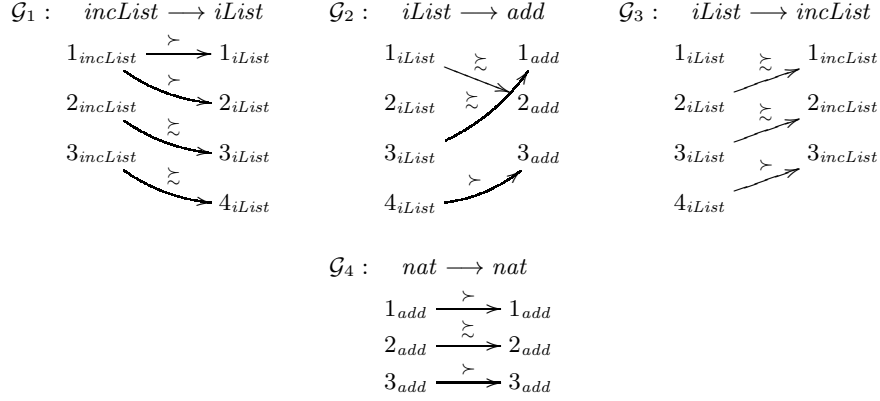
\begin{figure}[t]
  \centering
  $
  \begin{array}{ll@{~~~~~~}ll@{~~~~~~}ll}
  \cG_1: & {~~~\mathit{incList} \longrightarrow \mathit{iList}} 
    & \cG_2: & {~~~\mathit{iList} \longrightarrow \mathit{add}}
    & \cG_3: & {~~~\mathit{iList} \longrightarrow \mathit{incList}} \\
  & \xymatrix@R=6pt{
    1_{\mathit{incList}} \ar[r]^{\succ} \ar@/_/[dr]^{\succ} & 1_{\mathit{iList}} \\
    2_{\mathit{incList}} \ar@/_/[dr]^{\succsim} & 2_{\mathit{iList}} \\
    3_{\mathit{incList}} \ar@/_/[dr]^{\succsim} & 3_{\mathit{iList}} \\
    & 4_{\mathit{iList}} \\
  } & & \xymatrix@R=6pt{
    1_{\mathit{iList}} \ar[dr]^{\succsim} & 1_{\mathit{add}} \\
    2_{\mathit{iList}} & 2_{\mathit{add}} \\
    3_{\mathit{iList}} \ar@/_/[uur]^{\succsim} & 3_{\mathit{add}}\\ 
    4_{\mathit{iList}} \ar@/_/[ur]^{\succ} \\
  } & & \xymatrix@R=6pt{
    1_{\mathit{iList}} & 1_{\mathit{incList}} \\
    2_{\mathit{iList}} \ar[ur]^{\succsim} & 2_{\mathit{incList}} \\
    3_{\mathit{iList}} \ar[ur]^{\succsim} & 3_{\mathit{incList}} \\
    4_{\mathit{iList}} \ar[ur]^{\succ}
  } 
  \\[18ex]
   & & \cG_4: & {~~~\mathit{nat} \longrightarrow \mathit{nat}} & & \\
  & & & \xymatrix@R=1pt{
    1_{\mathit{add}} \ar[r]^{\succ} & 1_{\mathit{add}} \\
    2_{\mathit{add}} \ar[r]^{\succsim} & 2_{\mathit{add}} \\
    3_{\mathit{add}} \ar[r]^{\succ} & 3_{\mathit{add}} \\
  } &  & 
  \end{array}
  $
  \caption{Size-change graphs for $\mathit{incList}$}
  \label{graphs}
  \end{figure}

\begin{example} \label{ex1}
Consider the following simple program:
\[
\begin{array}{l@{~~~}l}
(c_1) & incList([\:],\_,[\:]).\\
(c_2) & incList([X|R],I,L) \leftarrow iList(X,R,I,L).\\
(c_3) & iList(X,R,I,[XI|RI]) \leftarrow add(I,X,XI), incList(R,I,RI).\\
(c_4) & add(0,Y,Y). \\
(c_5) & add(s(X),Y,s(Z)) \leftarrow add(X,Y,Z).
\end{array}
\]
Let $(\succsim,\succ)$ be the reduction pair induced by the symbolic
term-size norm $\snorm_{ts}$.  Here, we have four size-change graphs,
depicted in Fig.~\ref{graphs}, which are associated to clauses $c_2$
(graph $\cG_{1}$), $c_3$ (graphs $\cG_{2}$ and $\cG_3$) and $c_5$
(graph $\cG_4$).

\end{example}
In order to identify the program \emph{loops}, we should compute
roughly a transitive closure of the size-change graphs by composing
them in all possible ways. Basically, given two size-change graphs:
\[
\cG = (\{1_p,\ldots,n_p\}, \{1_q,\ldots,m_q\}, E_1) \hspace{5ex} \cH =
(\{1_q,\ldots,m_q\}, \{1_r,\ldots,l_r\}, E_2)
\]
w.r.t.\ the same reduction pair $(\succsim,\succ)$, their
concatenation is defined by
\[
\cG \conc \cH = (\{1_p,\ldots,n_p\},\{1_r,\ldots,l_r\},E)
\]
where $E$ contains an edge from $i_p$ to $k_r$ iff $E_1$ contains an
edge from $i_p$ to some $j_q$ and $E_2$ contains an edge from $j_q$ to
$k_r$. Furthermore, if some of the edges are labeled with $\succ$,
then so is the edge in $E$; otherwise, it is labeled with $\succsim$.

In particular, according to \cite{LJB01}, we only need to consider the
\emph{idempotent} size-change graphs $\cG$ with $\cG\conc\cG = \cG$
for analyzing the termination of the program.

\begin{example} \label{ex2}
  For the program of Example~\ref{ex1}, we compute the following
  idempotent size-change graphs:
\[
    \begin{array}{ll@{~~~~~~~~}ll@{~~~~~~~~}ll}
       & {incList ~~\longrightarrow~~ incList} 
      &  & {iList ~~\longrightarrow~~ iList} 
      &  & {add ~~\longrightarrow~~ add} \\
      & \xymatrix@R=0pt@C=20pt{
        1_{incList} \ar[r]^{\succ_{ts}} & 1_{incList} \\
        2_{incList} \ar[r]^{\succsim_{ts}} & 2_{incList} \\
        3_{incList} \ar[r]^{\succ_{ts}} & 3_{incList} \\
      }   & &
      \xymatrix@R=0pt@C=20pt{
        1_{iList} & 1_{iList} \\
        2_{iList} \ar[r]^{\succ_{ts}} \ar@/^/[ru]^{\succ_{ts}} & 2_{iList} \\
        3_{iList} \ar[r]^{\succsim_{ts}} & 3_{iList} \\
        4_{iList} \ar[r]^{\succ_{ts}} & 4_{iList} \\
      }   & &
      \xymatrix@R=0pt@C=20pt{
        1_{add} \ar[r]^{\succ_{ts}} & 1_{add} \\
        2_{add} \ar[r]^{\succsim_{ts}} & 2_{add} \\
        3_{add} \ar[r]^{\succ_{ts}} & 3_{add} \\
      }\end{array}
\]
that represent how the size of the arguments of the three potentially
looping predicates changes from one call to another.
\end{example}
Once the idempotent size-change graphs of a program have been
computed, the following results hold:\footnote{A term $t$ is
  \emph{instantiated enough} \cite{DLSS01,LS97} w.r.t.\ a symbolic
  norm $\snorm$ if $||t||$ is an integer constant.}
\begin{description}
\item[Termination:] An atom $A$ is (strongly) terminating if every
  idempotent size-change graph for $p/n$ contains at least one edge
  $i_p \stackrel{\succ}{\longrightarrow} i_p$ such that, for every
  computation rule $\cR$ and atom $p(t_1,\ldots,t_n) \in
  calls_P^\cR(A)$, the argument $t_i$ is instantiated enough w.r.t.\
  the considered symbolic norm.

  Clearly, the set $calls_P^\cR(A)$ is often infinite. Therefore, we
  usually consider an approximation based on a \emph{division} that
  classifies every predicate's argument as either \textsf{static} or
  \textsf{dynamic} and check that the $i$-th argument of $p$ is
  classified as \textsf{static} (rather than checking that $t_i$ is
  instantiated enough in all possible calls from $A$).

  For instance, given a division that classifies the arguments of
  $add$ as follows:
  \[
  add ~\mapsto~(\mathsf{static},\mathsf{dynamic},\mathsf{dynamic})
  \]
  and according to the idempotent size-change graphs of
  Example~\ref{ex2}, we have that all calls to $add$ terminate since
  there is an edge $1_{add} \stackrel{\succ}{\longrightarrow} 1_{add}$
  in the idempotent size-change graph and the first argument of $add$
  is classified as \textsf{static}.

\item[Quasi-termination:] An atom $A$ is (strongly) quasi-terminating
  if it is either terminating or every idempotent size-change graph
  for $p/n$ contains, for all $i_p$ ($i=1,\ldots,n$) an edge $j_p
  \stackrel{R}{\longrightarrow} i_p$ for some $j_p$, with
  $R\in\{\succ,\succsim\}$ (i.e., all arguments are bounded by the
  value of some argument in a previous call).
  Furthermore, the considered norms must be \emph{bounded} (see
  Definition~\ref{bnorm} below).

  For instance, according to the idempotent size-change graphs of
  Example~\ref{ex2}, an atom $add(X,Y,Z)$ is quasi-terminating since
  there is an input edge to every argument.
\end{description}
In \cite{Vid07}, the termination condition is used for ensuring the
local termination of partial evaluation, while the quasi-termination
condition is used for ensuring its global termination. Basically,
\begin{itemize}
\item we reclassify as \textsf{unfold} those atoms which are
  terminating w.r.t.\ a given division (and with \textsf{memo}
  otherwise) and

\item we mark with \textsf{dynamic} the argument of an atom if there
  is no input edge to this argument in some idempotent size-change
  graph, i.e., if the atom is not quasi-terminating.
\end{itemize}

\begin{example} \label{ex3}
  Given the idempotent size-change graphs of Example~\ref{ex2} and a
  division that classifies the predicates' arguments as follows:
  \[
  \begin{array}{lcl}
    \mathit{incList} ~\mapsto~ (\mathsf{dynamic},\mathsf{static},\mathsf{dynamic})\\
    \mathit{iList} ~\mapsto~ (\mathsf{dynamic},\mathsf{dynamic},\mathsf{static},\mathsf{dynamic})\\
    \mathit{add} ~\mapsto~ (\mathsf{static},\mathsf{dynamic},\mathsf{dynamic})\\
  \end{array}
  \]
  we have that
  \begin{itemize}
  \item $\mathit{incList}$ and $\mathit{iList}$ are marked with
    \textsf{memo} while $\mathit{add}$ is marked with \textsf{unfold},
    and
  \item no argument should be re-classified as \textsf{dynamic}.
  \end{itemize}
\end{example}

\section{Improving Size-Change Analysis}

In this section, we introduce several extensions of the size-change
analysis that may improve the accuracy of the specialization process
by taking into account some basic properties of partial evaluation.

\subsection{Non-Bounded Norms for Global Termination}

Let us recall the notion of \emph{bounded} norm required in
\cite{Vid07} for ensuring quasi-termination:

\begin{definition}[bounded norm] \label{bnorm}
  We say that a symbolic norm $\snorm$ is bounded if the set $\{ s
  \mid ||t||\sgeq||s||\}$ contains a finite number of nonvariant terms
  for any term $t$.
\end{definition}
Roughly speaking, a symbolic norm is bounded if, for every term $t$,
there exist only finitely many nonvariant terms whose weights are
lesser than or equal to that of $t$ w.r.t.\ the symbolic norm
$\snorm$.

Unfortunately, many symbolic norms are not bounded; e.g., the symbolic
list-length norm is not bounded since, given the term $p([a])$, we
have an infinite set $\{p([a]),p([f(a)]),p([f(f(a))]),\ldots\}$ of
non-variant terms such that $||[a]||_{ll} = ||[f(a)]||_{ll} = ||
[f(f(a))]||_{ll} = \ldots = 1$.

In the context of partial evaluation, however, symbolic norms need not
be bounded if the \emph{problematic} parts of the terms are
generalized at the global level. For instance, we can safely use the
symbolic list-length norm as long as the list elements are replaced by
fresh variables in the global level. This idea, already sketched in
\cite{LV08}, is formalized by means of the \emph{most general
  generalization} operator:

\begin{definition}[$mgg$]
  Let $\snorm$ be a symbolic norm. Given a term $t$, we denote by
  $\mgg^{\snorm}(t)$ the most general generalization of $t$ such that
  $||t|| = ||\mgg^{\snorm}(t)||$.  
  We also let $\mgg^{\snorm}(p(t_1,\ldots,t_n)) =
  p(\mgg^{\snorm}(t_1),\ldots,\mgg^{\snorm}(t_n))$.
\end{definition}
For instance, given the term $t = [s(N),b]$, we have
$\mgg^{\snorm_{ll}}(t) = [X,Y]$ but $\mgg^{\snorm_{ts}}(t) =
[s(N),b]$.

Moreover, the quasi-termination result in \cite{Vid07} also requires
that all calls encountered during partial evaluation should be
\emph{linear} w.r.t.\ the dynamic variables (i.e., no variable marked
as dynamic could appear more than once in a call). However, this is
not a real problem in the context of partial evaluation since all
dynamic parts of terms are replaced by fresh variables in the global
level anyway.

Therefore, one can ensure the global termination of partial evaluation
when using arbitrary symbolic norms in the size-change analysis as
long as
\begin{itemize}
\item dynamic parts of arguments are replaced by fresh variables in
  the global level (this is already done by current offline partial
  evaluators) and

\item an atom $A$ is replaced by $mgg^{\snorm}(A)$ in the global
  level, where $\snorm$ is the symbolic norm used in the size-change
  analysis.
\end{itemize}

\subsection{Maximizing ``Unfold'' Annotations}

The original approach of \cite{Vid07} does not take into account that
different idempotent size-change graphs may represent a single loop.
For instance, the idempotent size-change graphs for both
$\mathit{incList}$ and $\mathit{iList}$ actually represent the same
program loop. Therefore, it would be safe to annotate only one of
these predicates with ``\textsf{memo}'' and the other one with
``\textsf{unfold}''.

In order to avoid unnecessary \textsf{memo} annotations, one can
slightly extend the original annotation procedure as follows:
\begin{itemize}
\item First, every size-change graph is labeled with a unique
  identifier (e.g., $\cG_1$, $\cG_2$, \ldots, as in
  Fig.~\ref{graphs}).

\item Then, the concatenation of graphs is performed as before, but
  now every concatenation keeps a set with the identifiers of the
  graphs involved in the concatenation. We note that the set of
  identifiers is not taken into account during the concatenation
  process, i.e., two size-change graphs that only differ in the
  associated set of identifiers are considered equal (therefore, the
  complexity of the concatenation process, the most expensive part of
  the analysis, remains the same).

  For instance, the labeled idempotent size-change graphs of
  Example~\ref{ex2} would now be as depicted in Fig.~\ref{lgraphs}.

  \begin{figure}[t]
  \[
    \begin{array}{ll@{~~~~}ll@{~~~~}ll}
      \{\cG_1,\cG_3\}: & {incList ~~\longrightarrow~~ incList} 
      &  \{\cG_1,\cG_3\}: & {iList ~~\longrightarrow~~ iList} 
      & \{\cG_4\}: & {add ~~\longrightarrow~~ add} \\
      & \xymatrix@R=0pt@C=20pt{
        1_{incList} \ar[r]^{\succ_{ts}} & 1_{incList} \\
        2_{incList} \ar[r]^{\succsim_{ts}} & 2_{incList} \\
        3_{incList} \ar[r]^{\succ_{ts}} & 3_{incList} \\
      }   & &
      \xymatrix@R=0pt@C=20pt{
        1_{iList} & 1_{iList} \\
        2_{iList} \ar[r]^{\succ_{ts}} \ar@/^/[ru]^{\succ_{ts}} & 2_{iList} \\
        3_{iList} \ar[r]^{\succsim_{ts}} & 3_{iList} \\
        4_{iList} \ar[r]^{\succ_{ts}} & 4_{iList} \\
      }   & &
      \xymatrix@R=0pt@C=20pt{
        1_{add} \ar[r]^{\succ_{ts}} & 1_{add} \\
        2_{add} \ar[r]^{\succsim_{ts}} & 2_{add} \\
        3_{add} \ar[r]^{\succ_{ts}} & 3_{add} \\
      }\end{array}
    \]
    \caption{Labeled idempotent size-change graphs for
      $\mathit{incList}$}
    \label{lgraphs}
    \end{figure}
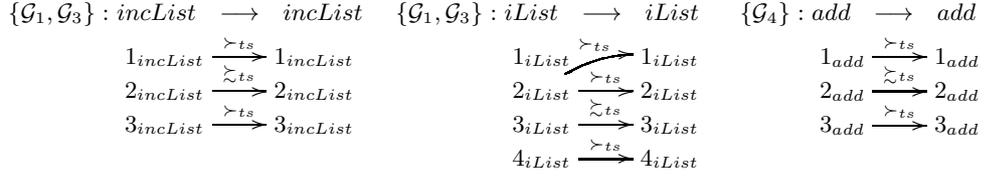

  \item The computed idempotent size-change graphs can now be grouped
    into equivalence classes so that two idempotent size-change graphs
    belong to the same class if they are labeled with the same set of
    identifiers.

  \item Finally, we should only annotate with ``\textsf{memo}'' one
    predicate for every equivalence class of idempotent size-change
    graphs.
\end{itemize}
For instance, as mentioned in Example~\ref{ex3}, both
$\mathit{incList}$ and $\mathit{iList}$ are marked with \textsf{memo}
in the original framework. Now, however, only one of them would be
marked with \textsf{memo} (and the other one with \textsf{unfold}).

Clearly, there is a degree of freedom when choosing which is the
idempotent size-change graph of a given class that should be marked
with \textsf{memo}.  For this purpose, one can define appropriate
heuristics that minimize the number of \textsf{memo} annotations by,
e.g., assigning a higher priority to those predicates that belong to
more than one class.

\subsection{Right-Propagation of Bindings}

An advantage of the size-change analysis of \cite{Vid07} is that it is
independent of a particular selection rule. As mentioned in the
introduction, this property makes the associated binding-time analysis
much faster; unfortunately, it is also less accurate.

In some cases, we can improve this situation by assuming some partial
knowledge on the evaluation order.\footnote{We thank Maurice
  Bruynooghe for suggesting this improvement.} For instance, we could
first run a left-termination analysis (like, e.g., the one based on
the binary unfoldings \cite{CT99}) or rely on user's annotations that
identify some atoms as ``completely unfoldable'' (note that an
annotation \textsf{unfold} only means that the atom can be unfolded
\emph{one step}; then the annotations of the predicates in the
unfolded goal should be followed).

In this case, we can improve the accuracy of the size-change analysis
by using an inter-argument size analysis like that calculated from the
convex hull of \cite{BKM05}. For instance, given the program
\[
\begin{array}{l}
p(X) ~\leftarrow~q(X,Y),p(Y). \\[1ex]
q(s(0),0).\\
q(s(X),Y) ~\leftarrow~ q(X,Y).
\end{array}
\]
the size-change graph associated to $p/1$ originally contains no edge
(since we do not know the size relation between $X$ and $Y$). Now, if
we assume that $q/2$ is completely unfoldable, then we can use the
output of the convex hull of \cite{BKM05} (using a term-size norm):
\[
  q(A,B) ~\leftarrow~ \{A>B,~B=0,~A\geq 1\}
\]
for propagating some additional constraints to the right of $q$. In
this way, one can easily infer that the size-change graph for $p/1$
should contain an edge $1_p \stackrel{\succ}{\longrightarrow} 1_p$.

Let us note that, in principle, the accuracy of the size-change
analysis of \cite{Vid07} could not be improved by adding
inter-argument size relations to size-change graphs, since
inter-argument relations usually require the atoms to be completely
unfolded (i.e., they represent relations that hold for success
patterns). This assumption is not generally true in the setting of
\cite{Vid07} where partial evaluations are possible.

\section{Discussion}

We have recently undertaken the implementation of a binding-time
analysis for the offline partial evaluation of Prolog programs which
is based on the size-change analysis of \cite{Vid07}. In this paper,
we have introduced several improvements that may allow us to overcome
the main weaknesses of \cite{Vid07}. An experimental evaluation will
be conducted in order to assess their effectiveness in practice.


\end{document}